\documentclass[a4paper,twoside,twocolumn,english,prb,showpacs]{revtex4}
\usepackage[T1]{fontenc}
\usepackage[latin1]{inputenc}
\usepackage{longtable}
\usepackage{graphicx}
\usepackage{amssymb}

\makeatletter

\providecommand{\tabularnewline}{\\}

\usepackage{babel}
\makeatother
\begin{document}

\title{Local dissipation effects in two-dimensional quantum Josephson junction
arrays with magnetic field }

\author{T. P. Polak}

\address{Max-Planck-Institut f\"ur Physik komplexer Systeme, N\"othnitzer
Straße 38, 01187 Dresden, Germany}

\author{T. K. Kope\'c}

\address{Institute for Low Temperatures and Structure Research, Polish Academy
of Sciences, POB 1410, 50-950 Wroclaw 2, Poland}


\begin{abstract}
We study the quantum phase transitions in two-dimensional arrays of
Josephson-couples junctions with short range Josephson couplings (given
by the Josephson energy $E_{J}$) and the charging energy $E_{C}$.
We map the problem onto the solvable quantum generalization of the
spherical model that improves over the mean-field theory method. The
arrays are placed on the top of a two-dimensional electron gas separated
by an insulator. We include effects of the local dissipation in the
presence of an external magnetic flux $f=\Phi/\Phi_{0}$ in square
lattice for several rational fluxes $f=0,\frac{1}{2},\frac{1}{3},\frac{1}{4}$
and $\frac{1}{6}$. We also have examined the $T=0$ superconducting-insulator
phase boundary as function of a dissipation $\alpha_{0}$ for two
different geometry of the lattice: square and triangular. We have
found critical value of the dissipation parameter independent on geometry
of the lattice and presence magnetic field.
\end{abstract}
\maketitle

\section{Introduction}

Over the past several years Josephson junction arrays (JJA) have gained
increased interest since the display quantum mechanics on macroscopic
scale. In the last years, due to development of the micro-fabrication
techniques, it became possible to fabricate Josephson junction arrays
with tailor specific properties. In these systems the superconductor-insulator
(SI) transition can be driven by quantum fluctuations when the charging
energy $E_{C}$ becomes comparable to the Josephson coupling energy
$E_{J}$.\cite{abeles,simanek1,simanek2,doniach,bradley,kopec1,wood,jose}
It has been established that in arrays which are in the superconducting
state, but with value $E_{C}/E_{J}$ close to critical value, a magnetic
field can be used to drive array into the insulating state.\cite{kopec2}
There are experimental possibilities to fabricate different structures
of the JJA like square\cite{voss,wess,zant} and triangular\cite{zant}
networks which show that the critical value $E_{J}/E_{C}$ can change
its value depending on the geometry of the lattice. Zant et al. showed
experimentally that not only geometry can influence on value $E_{C}/E_{J}$
in JJA but also external magnetic field can lead system to the phase
transition.\cite{zant,caputo} In a JJA an applied gate voltage relative
to the ground plane $V_{g}$ introduces a charge frustration.\cite{bruder}
The combination of charge frustration and Coulomb interaction results
in the appearance of various Mott insulating phases separated by regions
of phase coherent superconducting state.

It has been understood early that dissipation can be capable of driving
an SI transitions. Problem how dissipation can be described on quantum
mechanical level was firstly addressed by Caldeira and Leggett.\cite{caldeira}
In this formalism dissipation introduces damping of phase fluctuations
that is inversely proportional to the resistance of the environment.
Quantum phase model of JJA was formulated in terms of an effective
action.\cite{schon,ambegaokar} Various theoretical methods have been
applied in effects of Ohmic dissipation\cite{chakravaraty,fisher,simanek3,fisher1}
such as coarse graining,\cite{zwerger,panyukov} variational\cite{panyukov,chakravaraty,chakravarty1,falci,cuccoli,chakravarty2}
and renormalization group approaches.\cite{panyukov,chakravarty1,fisher,fisher1}
The dissipation due to quasi-particle tunneling in JJA\cite{eckern}
was investigated by means of the mean field calculations,\cite{chakravarty1}
variational approaches\cite{kampf1} and Monte Carlo simulations.\cite{choi}
Phase transitions of dissipative JJA's in magnetic field were analyzed
by Kampf and Sch\"on\cite{kampf} and relied upon mean-field approximation
mapped into tight-binding Schrödinger equation for Bloch electrons
in magnetic field. It has been shown that at a fixed temperature we
can observe a phase transition if we vary the magnetic field. The
similar problem was solved by Kim and Choi\cite{kim} based on effective
Hamiltonian obtained by the Feynman path integral formalism. Authors
claim that especially in low temperatures variational method is not
precise enough to perceive such a subtle transition. 

Dissipation may arise from coupling with the substrate by means of
the local damping model.\cite{beck} These studies indeed revealed
the existence of a critical value of the sheet resistance which seems
to agree with experimental results in JJA\cite{geerlings} and thin
films.\cite{valles} On the other hand some experimental studies discloses
that the values of the critical resistance can show wide variations
quite the contrary to the predicted universal value close to $h/4e^{2}$.\cite{yazdani}
Furthermore, previous theoretical calculations relied upon variational
and mean field approximations, which usually are not expected to be
reliable at $T=0$ and be capable to handle spatial and quantum fluctuations
effects properly, especially in two dimensions.

It has been shown\cite{fazio} that there is a possibility of existence
four phases in superconducting junction arrays with dissipation and
the phase diagram depends on ratio $E_{J}/E_{C}$ and dissipation
parameter $\alpha_{0}$. One is insulating, when both Cooper pairs
and single electrons are frozen ($E_{J}/E_{C}$ and $\alpha_{0}$
small), mixed, when superconducting long-range coherence and single
electrons tunneling coexist ($E_{J}/E_{C}$ and $\alpha_{0}$ large),
and for intermediate values of the parameters we can obtain ($E_{J}/E_{C}$
small and $\alpha_{0}$ large) Cooper pairs remain frozen, but single
electrons are free. In the opposite limit ($E_{J}/E_{C}$ large and
$\alpha_{0}$ small) we have superconducting long-range phase order
and there is not any dynamics of the single electrons.

Fact, that dissipation could play important role in solid state physics
appeared recently from the high-$T_{\mathrm{C}}$ point of view.  Relation
obtained by Homes et all.\cite{homes} proves that the characteristic
timescale for dissipation could not be shorter than in the high-$T_{\mathrm{C}}$
superconductors. Relaxation time $\tau_{diss}\sim\hbar/k_{\mathbf{\mathrm{B}}}T_{C}$
(Planckian dissipation) is an essence of the Home's law and there
is evidence that, quantum-critical nature of the system could be present
even in the normal state. It means that below this timescale we have
purely quantum mechanical behaviour and energy could not be turned
out into heat. Conductivity in the normal state is tied with the relaxation
time relation $\sigma_{normal}\sim\tau_{diss}$ and it indicates that
$\sigma_{normal}$ should be as small as it is allowed by Planck's
constant.

Realization of a quantum computer crucially depends on our ability
to create and hold coherent superposition states, so decoherence presents
the most fundamental trammel. Especially coupling between different
devices and environment achieves to dissipation, and hence decoherence
which leads to exponential decay of superposition states into incoherent
mixtures.\cite{zurek} Both in superconducting qubits, based on superconducting
interference devices and in single-pair tunneling devices the Josephson
junction is a key element and it is the dissipation of the junction
that sets the limit on decoherence time.\cite{han}

To purpose of this paper is to study local dissipation effects in
JJA in an analytical way to refine the phase diagram of the system
for different geometry of the lattices and in presence of the external
magnetic field. We consider a network of the Josephson junction arrays
with tunable dissipation which is placed on the top of a two-dimensional
electron gas (2DEG) separated by an insulator. We drop nonlocal charging
and dissipative terms. The problem we would like to address is then:
What is the effect of the competition between magnetic, geometric
and quantum effects on the ground state ordering in the two-dimensional
Josephson arrays? To analyze 2D JJA beyond mean field approximation
we employ the path-integral formulation of quantum mechanics and a
quantum spherical model approach accurately tailored for the JJA systems.
This formalism allows then for explicit implementation of the local
dissipation effects and magnetic field into ours considerations. Most
theoretical studies investigated the simple square lattice geometry
of JJA. Another structures were analyzed by Monte Carlo simulations
and mean field calculations in magnetic field in the context of phase
transitions\cite{shih}. On the other hand Granato and Kosterlitz
claim that transition in 2D array with differential geometry can be
described by classical Ginzburg-Landau-Wilson effective free energy
with two complex field.\cite{granato} We analyze the quantitative
changes in the phase diagram due to two different geometrical JJA
structures without external magnetic field. We consider influence
of the magnetic field on square lattice because many different properties
of an array depend on flux parameter $f=p/q$.\cite{zant,benz}

The paper is organized as follows. In the next section we introduce
the model. In Sec. IIc we formulate the problem in terms of the effective
action of quantum spherical model. The various zero temperature phase
diagrams are then studied in Sec III. Finally, in Sec IV we summarize.

\section{Model}

\begin{figure}
\includegraphics[%
  scale=0.38]{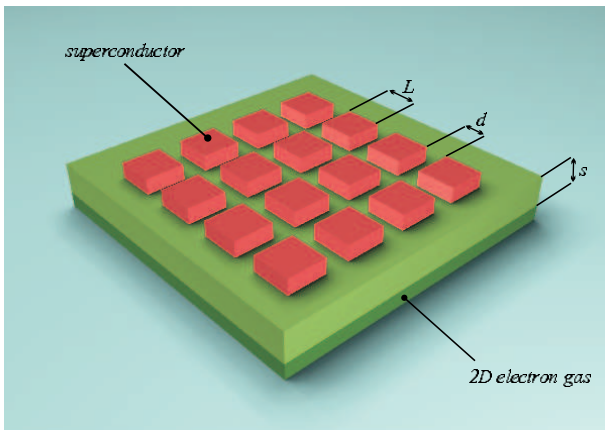}

\caption{Schematic view of the 2D JJA in dissipative environment.\label{Fig1}}
\end{figure}

We consider a two-dimensional Josephson junction array with lattice
sites $i$, characterized by superconducting phase $\phi_{i}$ in
dissipative environment. Possible experimental realization of the
2D JJA is shown on the Fig.\ref{Fig1}. The array can be formed by
thin square superconducting islands of size $L$. The separation $d$
between islands must be small enough to allow for the Josephson interactions.
The variable dissipation is introduced by coupling with two-dimensional
electron gas (Ohmic bath) which is localized within distance $s$.
The corresponding Euclidean action in the Matsubara imaginary time
$\tau$ formulation $\left(0\leq\tau\leq1/k_{\mathrm{B}}T\equiv\beta\right)$,
with $T$ being temperature and $k_{\mathrm{B}}$ the Boltzmann constant
($\hbar=1$)

\begin{equation}
\mathcal{S}=\mathcal{S}_{\mathrm{C}}+\mathcal{S}_{\mathrm{J}}+\mathcal{S}_{\mathrm{D}},\label{sum action}\end{equation}
where\begin{eqnarray}
\mathcal{S}_{\mathrm{C}} & = & \frac{1}{16E_{C}}\sum_{i}\int_{0}^{\beta}d\tau\left(\frac{d\phi_{i}}{d\tau}\right)^{2},\nonumber \\
\mathcal{S}_{\mathrm{J}} & = & \sum_{\left\langle i,j\right\rangle }\int_{0}^{\beta}d\tau J_{ij}\left\{ 1-\cos\left[\phi_{i}\left(\tau\right)-\phi_{j}\left(\tau\right)\right]\right\} ,\nonumber \\
\mathcal{S}_{\mathrm{D}} & = & \frac{1}{2}\sum_{i}\int_{0}^{\beta}d\tau d\tau'\alpha\left(\tau-\tau'\right)\left[\phi_{i}\left(\tau\right)-\phi_{i}\left(\tau'\right)\right]^{2}.\label{action}\end{eqnarray}
 The first part of the action Eq. (\ref{action}) defines the electrostatic
energy, where self charging energy parameter is \begin{equation}
E_{C}=\frac{e^{2}}{2C_{0}}.\label{culomb energy}\end{equation}
The second term is the Josephson energy $E_{J}$ ($J_{ij}\equiv E_{J}$
for $\left|i-j\right|=\left|d\right|$ and zero otherwise). Third
part of the action $\mathcal{S}_{\mathrm{D}}$ describes the local
dissipation where $\alpha\left(\tau-\tau'\right)$ is dissipative
kernel. For a local dissipation effects Fourier transform (with respect
to imaginary times) of the kernel Eq. (\ref{action}) is \cite{ambegaokar,eckern}
\begin{equation}
\alpha\left(\omega_{n}\right)=\frac{\alpha_{0}}{\pi}\left|\omega_{n}\right|,\label{dissipation kernel}\end{equation}
where dimensionless parameter: \begin{equation}
\alpha_{0}=\frac{R_{Q}}{R_{0}},\qquad R_{Q}=\frac{1}{4e^{2}}.\end{equation}
describes strength of the Ohmic dissipation. Here the $R_{0}$ is
the shunt to the ground and is interpreted as the shunt resistance
present in many experiments. As we can see the dissipation part of
the action Eq. (\ref{action}) breaks the $2\pi$-periodicity in the
phase variables since it allows for continuous charge fluctuations.
In proximity-coupled arrays, dissipation can correlate the phase of
a single island in different time.

\subsection{Effect of the applied magnetic field}

The perpendicular magnetic field $B$ given by vector potential $\mathbf{A}$
enters action Eq. (\ref{action}) through a Peierls factor according
to\begin{equation}
J_{ij}\rightarrow J_{ij}\exp\left(\frac{2\pi i}{\Phi_{0}}\int_{\mathbf{r}_{i}}^{\mathbf{r}_{j}}\mathbf{A}\cdot d\mathbf{l}\right),\end{equation}
 where $\Phi_{0}=2\pi c/e$ is a quantum of magnetic flux piercing
a 2D lattice plaquette and integral runs between center of grains
$\mathbf{r}_{i}$ and $\mathbf{r}_{j}$. Thus, the phase shift on
each junction is determined by the vector potential $\mathbf{A}$
and in typical experimental situation it can be entirely ascribed
to the external field $B$. The presence of $B$ induces vortices
in the system described by the flux parameter $f$($\equiv\Phi/\Phi_{0}=Ba^{2}/\Phi_{0}$
where $a$ is area of an elementary plaquette). Our special interest
are the values of the magnetic field when flux parameter $f=p/q$
represents rational values. %

\subsection{Method }

To study the JJA model we can use a description in terms of an effective
Ginzburg-Landau functional derived from the microscopic model Eq.
(\ref{action}). Several studies of JJA have followed this way, also
known as coarse grained approach first developed by Doniach.\cite{doniach}
Most of existing analytical works on quantum JJA have employed different
kinds of mean-field-like approximations which are not reliable for
treatment spatial and temporal quantum phase fluctuations which quantum
spherical approach captures. Formulation of the problem in terms of
the spherical model\cite{berlin,joyce} which was initiated by Kope\'c
and Jos\'e\cite{kopec3} leads us to introduce the auxiliary complex
field $\psi_{i}$ whose expectation value is proportional to the $\mathbf{S}_{i}\left(\phi\right)$
vector defined by

\begin{equation}
\mathbf{S}_{i}\left(\phi\right)=\left[S_{i}^{x}\left(\phi\right),S_{i}^{y}\left(\phi\right)\right]\equiv\left[\cos\left(\phi_{i}\right),\sin\left(\phi_{i}\right)\right].\end{equation}
Natural consequence definition $\mathbf{S}_{i}\left(\phi\right)$
is the following rigid constraint\begin{equation}
\left|\mathbf{S}_{i}\left(\phi\right)\right|^{2}=\left[S_{i}^{x}\left(\phi\right)\right]^{2}+\left[S_{i}^{y}\left(\phi\right)\right]^{2}=\cos^{2}\left(\phi_{i}\right)+\sin^{2}\left(\phi_{i}\right)\equiv1.\label{constraint}\end{equation}
The relation in Eq. (\ref{constraint}) implies that a weaker condition
also applies, namely:

\begin{equation}
\sum_{i}\left|\mathbf{S}_{i}\left(\phi\right)\right|^{2}=N.\label{spherical constraint}\end{equation}
Using the Fadeev-Popov method with the Dirac $\delta$ functional
we obtain:

\begin{eqnarray}
\mathcal{Z} & = & \int\left[\prod_{i}\mathcal{D}\psi_{i}\mathcal{D}\psi_{j}^{*}\right]\delta\left(\sum_{i}\left|\psi\left(\tau\right)\right|^{2}-N\right)e^{-\mathcal{S}_{\mathrm{J}}\left[\psi\right]}\nonumber \\
 &  & \times\int\left[\mathcal{D}\phi_{i}\right]e^{-\mathcal{S}_{\mathrm{C}+\mathrm{D}}\left[\phi\right]}\prod_{i}\delta\left[\Re\psi_{i}\left(\tau\right)-S_{i}^{x}\left(\phi\left(\tau\right)\right)\right]\nonumber \\
 &  & \times\delta\left[\Im\psi_{i}\left(\tau\right)-S_{i}^{y}\left(\phi\left(\tau\right)\right)\right].\end{eqnarray}
It is convenient to employ the functional Fourier representation of
the $\delta$ functional to enforce the spherical constraint Eq. (\ref{spherical constraint}):
\begin{equation}
\delta\left[x\left(\tau\right)\right]=\int_{-i\infty}^{+i\infty}\left[\frac{\mathcal{D}\lambda}{2\pi i}\right]\exp\left[\int_{0}^{\beta}d\tau\lambda\left(\tau\right)x\left(\tau\right)\right],\end{equation}
which introduces the Lagrange multiplier $\lambda\left(\tau\right)$
thus adding an quadratic term (in $\psi$ field) to the action Eq.
(\ref{action}). To evaluation of the effective action in terms of
the $\psi$ to second order in $\psi_{i}\left(\tau\right)$ gives
the partition function of the quantum spherical model\begin{eqnarray}
\mathcal{Z_{\mathrm{QSM}}} & = & \int\left[\prod_{i}\mathcal{D}\psi_{i}\mathcal{D}\psi_{j}^{*}\right]\nonumber \\
 &  & \times\delta\left(\sum_{i}\left|\psi\left(\tau\right)\right|^{2}-N\right)e^{-\mathcal{S}_{\mathrm{QSA}}\left[\psi\right]}\label{partition function}\end{eqnarray}
where the action of the effective nonlinear $\sigma$ model reads:\begin{eqnarray}
\mathcal{S}_{\mathrm{QSA}}\mathrm{\left[\psi,\lambda\right]} & = & \sum_{\left\langle i,j\right\rangle }\int_{0}^{\beta}d\tau d\tau'\left\{ \left[J_{ij}\left(\tau\right)\delta\left(\tau-\tau'\right)\right.\right.\nonumber \\
 & + & \left.\mathcal{W}_{ij}^{-1}\left(\tau,\tau'\right)-\lambda\left(\tau\right)\delta_{ij}\delta\left(\tau-\tau'\right)\right]\psi_{i}\psi_{j}^{*}\nonumber \\
 & + & \left.N\lambda\left(\tau\right)\delta\left(\tau-\tau'\right)\right\} .\label{sqsa}\end{eqnarray}
Furthermore 

\begin{eqnarray}
\mathcal{W}_{ij}\left(\tau,\tau'\right) & = & \frac{\delta_{ij}}{\mathcal{Z}_{0}}\int\left[\prod_{i}\mathcal{D}\phi_{i}\right]e^{i\left[\phi_{i}\left(\tau\right)-\phi_{j}\left(\tau'\right)\right]}e^{-\mathcal{S}_{\mathrm{C}+\mathrm{D}}\left[\phi\right]}\nonumber \\
 & \equiv & \mathcal{W}\left(\tau,\tau'\right)\delta_{ij},\label{correlator}\end{eqnarray}
is the phase-phase correlation function with statistical sum

\begin{equation}
\mathcal{Z}_{0}=\int\left[\prod_{i}\mathcal{D}\phi_{i}\right]e^{-\mathcal{S}_{\mathrm{C}+\mathrm{D}}\left[\phi\right]},\end{equation}
where action $\mathcal{S}_{\mathrm{C}+\mathrm{D}}\left[\phi\right]$
is just a sum electrostatic and dissipative term in Eq. (\ref{action}).
After introducing the Fourier transform of the field\begin{equation}
\phi_{i}\left(\tau\right)=\frac{1}{N\beta}\sum_{\mathbf{k}}\sum_{n=-\infty}^{+\infty}\phi_{\mathbf{k},n}\exp\left[-\left(i\omega_{n}\tau-\mathbf{k}\mathbf{r}_{i}\right)\right]\end{equation}
with $\omega_{n}=2\pi n/\beta$, $\left(n=0,\pm1,\pm2,...\right)$
being the Bose Matsubara frequencies, we can write expression Eq.
(\ref{correlator}) in form\begin{widetext}\begin{eqnarray}
\mathcal{W}\left(\tau,\tau'\right) & = & \frac{1}{\mathcal{Z}_{0}}\int\left[\prod_{\mathbf{k}}\mathcal{D}\phi_{\mathbf{k},n}\phi_{\mathbf{k},n}^{*}\right]\exp\left\{ -\frac{1}{4N\beta}\sum_{n=-\infty}^{+\infty}\left[\frac{1}{4E_{C}}\omega_{n}^{2}+\frac{\alpha_{0}}{\pi}\left|\omega_{n}\right|\right]\phi_{\mathbf{k},n}\phi_{\mathbf{k},n}^{*}+\right.\nonumber \\
 &  & \left.+\frac{1}{\beta N}\sum_{n=-\infty}^{+\infty}\left[\phi_{\mathbf{k},n}\left(e^{i\omega_{n}\tau}-e^{-i\omega_{n}\tau'}\right)+\phi_{\mathbf{k},n}^{*}\left(e^{-i\omega_{n}\tau}-e^{-i\omega_{n}\tau'}\right)\right]\right\} .\label{correlation function k tau}\end{eqnarray}
\end{widetext}Using Hubbard-Stratonovich transformation the phase-phase
correlation function reads:

\begin{equation}
\mathcal{W}\left(\tau,\tau'\right)=\exp\left\{ -\frac{1}{\beta}\sum_{n\neq0}\frac{1-\cos\left[\omega_{n}\left(\tau-\tau'\right)\right]}{\frac{1}{8E_{C}}\omega_{n}^{2}+\frac{\alpha_{0}}{2\pi}\left|\omega_{n}\right|}\right\} .\label{correlation function}\end{equation}
 The low temperature properties of the expression $\mathcal{W}^{-1}\left(\omega_{n}\right)$
lead to critical value $\alpha_{0}=2$ {[}see Appendix A{]}. Finally,
for small frequencies and $\alpha_{0}\leq2$ inverse of the correlation
function Eq. (\ref{correlation function}) becomes:\begin{equation}
\mathcal{W}^{-1}\left(\omega_{n}\right)=\frac{1}{8E_{C}}\omega_{n}^{2}+\frac{\alpha_{0}}{2\pi}\left|\omega_{n}\right|\end{equation}
for $\omega_{n}\neq0$ and $\mathcal{W}^{-1}\left(\omega_{n}\right)=0$
otherwise. In the thermodynamic limit ($N\rightarrow\infty$) the
steepest descent methods becomes exact; the condition that the integrand
in Eq. (\ref{partition function}) has a saddle point $\lambda\left(\tau\right)=\lambda_{0}$,
leads to an implicit equation for $\lambda_{0}$:\begin{equation}
1=\frac{1}{\beta N}\sum_{\mathbf{k}}\sum_{n\neq0}G\left(\mathbf{k},\omega_{n}\right),\label{spherical constraint 1}\end{equation}
where\begin{equation}
G^{-1}\left(\mathbf{k},\omega_{n}\right)=\lambda_{0}-J_{\mathbf{k}}+\frac{1}{8E_{C}}\omega_{n}^{2}+\frac{\alpha_{0}}{2\pi}\left|\omega_{n}\right|\label{lagrange coefficient}\end{equation}
with $J_{\mathbf{k}}$ as Fourier transform of the Josephson couplings
$J_{ij}$.

\section{Phase diagrams}

A Fourier transform of Eq. (\ref{sqsa}) in momentum and frequency
space enables one to write the spherical constraint Eq. (\ref{spherical constraint 1})
explicitly as

\begin{equation}
1=\frac{1}{\beta}\int_{-\infty}^{+\infty}d\xi\sum_{n\neq0}\frac{\rho\left(\xi\right)}{\lambda-\xi E_{J}+\frac{1}{8E_{C}}\omega_{n}^{2}+\frac{\alpha_{0}}{2\pi}\left|\omega_{n}\right|}.\label{spherical constraint 2}\end{equation}
 As is usual in a spherical model, the critical behavior and the phase
transitions boundary crucially depends on the spectrum given by density
of states $\rho\left(\xi\right)$ and is determined by the denominator
of the summand in the spherical constraint Eq. (\ref{spherical constraint 2}).
To proceed, it is desirable to introduce density of states for two
dimensional lattice in form\begin{equation}
\rho\left(\xi\right)=\frac{1}{N}\sum_{\mathbf{k}}\delta\left(\xi-\frac{J_{\mathbf{k}}}{E_{J}}\right).\label{dos}\end{equation}
where $J_{\mathbf{k}}$ is energy dispersion. Problem of computing
density of states $\rho_{p/q}^{\square}\left(\xi\right)$ for two-dimensional
square lattice with uniform magnetic flux per unit plaquette reduces
to the solution of Harper's equation\cite{harper}, e.g. relevant
for tight-binding electrons on 2D lattice\cite{hasegawa}. Analytical
results for density of states for square lattice were presented recently\cite{kopec2},
and based on analytically solving Harper's equation and receiving
dispersion relation $J_{\mathbf{k},p/q}^{\square}$. Influence of
the local dissipation effects will be considered on triangular lattice
without magnetic field. In that case we can use definition Eq. (\ref{dos})
but the dependence on the wave vector is different and could be described
by expression \begin{equation}
J_{\mathbf{k}}^{\bigtriangleup}=E_{J}\left[\cos\left(k_{x}\right)+2\cos\left(\frac{1}{2}k_{x}\right)\cos\left(\frac{\sqrt{3}}{2}k_{y}\right)\right].\end{equation}
Closed formulas for the density of states are placed in Appendix B.
The systems displays a critical point at $\lambda_{0}=G^{-1}\left(\mathbf{k}=0,\omega_{n}=0\right)\equiv\max\left[J\left(\mathbf{k}=0\right)\right]$.
This fixes the saddle point value of the Lagrange multiplier, $\lambda$
sticks to that value at criticality ($\lambda=\lambda_{0}$) and stays
constant in the whole low temperature phase. By substituting the value
of $\lambda_{0}$ to Eq. (\ref{spherical constraint 2}) and after
performing the summation over Matsubara frequencies, in $T\rightarrow0$
limit we obtain the following result:

\begin{eqnarray}
1 & = & \frac{1}{\pi}\int_{-\infty}^{+\infty}d\xi\frac{\rho\left(\xi\right)}{\sqrt{\left(\frac{\alpha_{0}}{2\pi}\right)^{2}-\frac{J_{0}-\xi E_{J}}{2E_{C}}}}\nonumber \\
 &  & \times\ln\left[\frac{\frac{\alpha_{0}}{2\pi}+\sqrt{\left(\frac{\alpha_{0}}{2\pi}\right)^{2}-\frac{J_{0}-\xi E_{J}}{2E_{C}}}}{\frac{\alpha_{0}}{2\pi}-\sqrt{\left(\frac{\alpha_{0}}{2\pi}\right)^{2}-\frac{J_{0}-\xi E_{J}}{2E_{C}}}}\right].\label{critical line}\end{eqnarray}
 It is easy to see that by specifying density of states $\rho\left(\xi\right)$
Eq. (\ref{dos}) with the Coulomb energy $E_{C}$ Eq. (\ref{culomb energy})
for given superconducting network without and in presence of the external
magnetic field, we are able to study the zero temperature JJA phase
diagram. The solutions and boundary cases above equation we will examine
in the next subsections.%
\begin{figure}
\includegraphics[%
  scale=0.55]{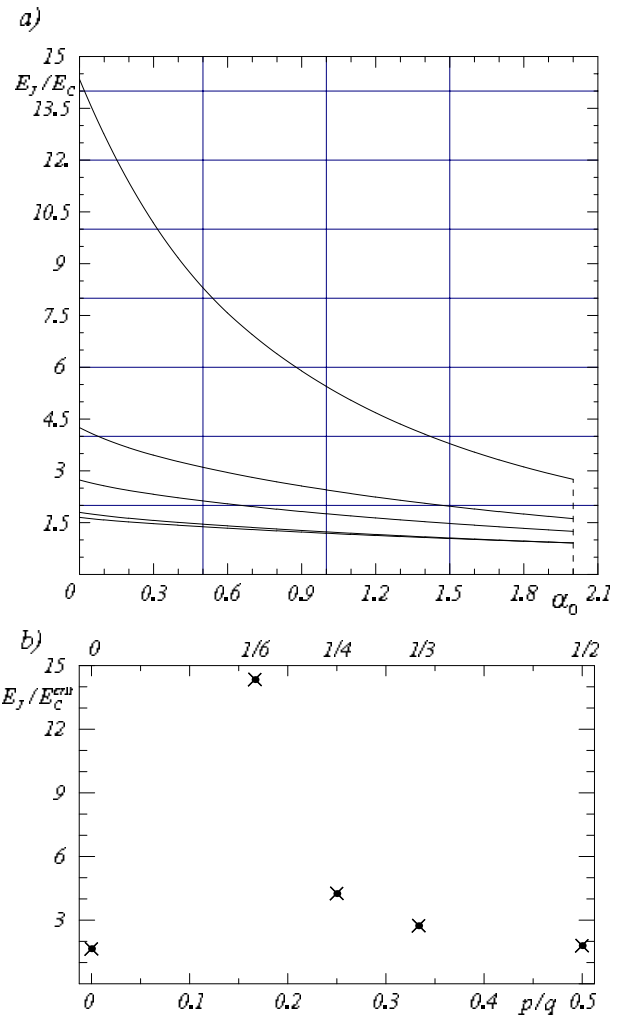}

\caption{a) Zero temperature phase diagram for the total charging energy $E_{J}/E_{C}$
vs parameter of dissipation $\alpha_{0}$ for square lattice. Insulating
state is below the curve. From the top we have $p/q=\frac{1}{6},\frac{1}{4},\frac{1}{3},\frac{1}{2}$
and $0$ b) Nonmonotonical dependence the critical value of the inverse
Coulomb energy $E_{J}/E_{C}^{crit}$ for several ratio of the magnetic
fluxes $p/q=0,\frac{1}{2},\frac{1}{3},\frac{1}{4}$ and $\frac{1}{6}$.\label{Fig2}}
\end{figure}

\subsection{Small $\alpha_{0}$ limit}

In the limit $\alpha_{0}\rightarrow0$ expression (\ref{critical line})
reduces to zero-temperature critical line in absence dissipation effects:\begin{equation}
1=\int_{-\infty}^{+\infty}d\xi\rho\left(\xi\right)\sqrt{\frac{2E_{C}}{J_{0}-\xi E_{J}}},\end{equation}
which is according with previous calculations\cite{kopec1}. When
$\alpha_{0}$ is nonzero, but still small, the situation changes due
to a dissipation is a factor which produces additional quantum fluctuations.
In consequence for small $\alpha_{0}$ the critical value $E_{J}/E_{C}$
decreases as:\begin{eqnarray}
E_{J}/E_{C} & = & A_{0}^{2}-\left(\frac{2}{\pi^{2}}\frac{A_{1}}{A_{0}}\right)^{2}\alpha^{2}+\frac{2}{A_{0}}\left(\frac{2}{\pi^{2}}\frac{A_{1}}{A_{0}}\right)^{3}\alpha^{3}+\nonumber \\
 &  & -\frac{5}{A_{0}^{2}}\left(\frac{2}{\pi^{2}}\frac{A_{1}}{A_{0}}\right)^{4}\alpha^{4}+\mathcal{O}\left(\alpha^{5}\right),\label{expansion small alpha}\end{eqnarray}
where corresponding coefficients reads: \begin{eqnarray}
A_{0} & = & \sqrt{2}\int_{-\infty}^{+\infty}d\xi\frac{\rho\left(\xi\right)}{\sqrt{J_{0}/E_{J}-\xi}},\label{factor1}\\
A_{1} & = & \int_{-\infty}^{+\infty}d\xi\frac{\rho\left(\xi\right)}{J_{0}/E_{J}-\xi}.\label{factor2}\end{eqnarray}

\begin{table}
\begin{longtable}{c|c|c|c|c|c|c}
\hline 
DOS&
$\rho^{\vartriangle}$&
$\rho_{0}^{\square}$&
$\rho_{1/2}^{\square}$&
$\rho_{1/3}^{\square}$&
$\rho_{1/4}^{\square}$&
$\rho_{1/6}^{\square}$\tabularnewline
\endhead
$A_{1}$&
$1.92619$&
$5.82281$&
$4.83852$&
$11.8065$&
$19.1268$&
$175.717$\tabularnewline
\hline
\endfoot
\hline 
$A_{0}$&
$1.01087$&
$1.28576$&
$1.34085$&
$1.65397$&
$2.06193$&
$3.78672$\tabularnewline
\end{longtable}

\caption{Factors of the expansion critical values $E_{J}/E_{C}$ for small
dissipation parameter $\alpha_{0}$ (Eq. \ref{factor1}, \ref{factor2}).\label{table1}}
\end{table}
The numerical values of the factors $A_{0}$ and $A_{1}$ are classified
in the Table (\ref{table1}). For small ratio $E_{J}/E_{C}$ there
is no chance to mobility both Cooper's pairs and single electrons.
If we increase the Josephson energy (or decrease Coulomb energy) we
induce phase transitions between insulating and superconducting phase.
We have global coherence state but in which there is no single electrons
dynamics. The critical values of the inverse Coulomb energy $E_{J}/E_{C}^{crit}$
for $\alpha_{0}=0$ are depicted in Fig.\ref{Fig2}b) for several
values of the magnetic field and are simply equal the first coefficient
of the expansion Eq. (\ref{expansion small alpha}): $\left.E_{J}/E_{C}^{crit}\right|_{\alpha_{0}=0}=A_{0}^{2}$.
We note the non-monotonic dependence of the Coulomb energy on the
magnetic flux ratio $p/q$.

\subsection{Critical value $\alpha_{0}$}

\begin{figure}
\includegraphics[%
  scale=0.45]{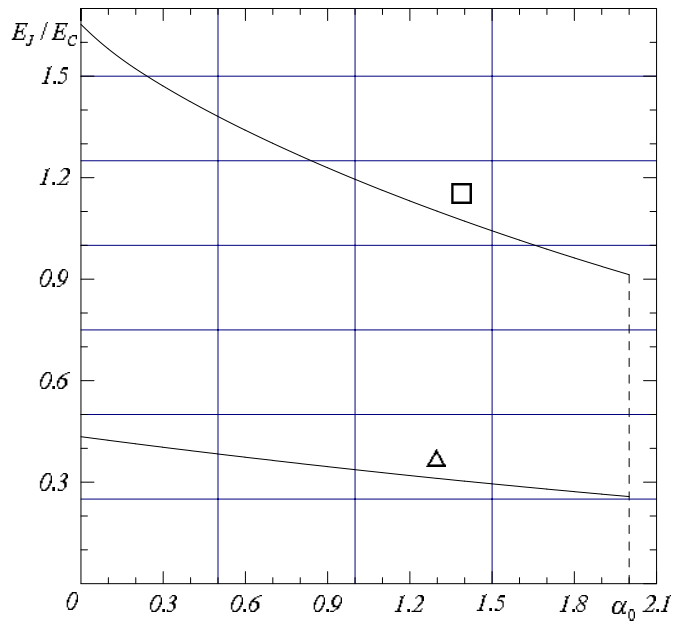}

\caption{Zero temperature phase diagram for the normalized Josephson energy
$E_{J}/E_{C}$ vs parameter of dissipation $\alpha_{0}$ for square
and triangular lattice. Insulating state is below the curves. \label{Fig3}}
\end{figure}

At zero temperature dissipation suppresses quantum fluctuations entirely
and drives system to a global coherent state. Due to the fact, that
correlation function Eq. (\ref{correlation function}) for low temperatures:
$\mathcal{W}^{-1}\left(\omega_{n}\right)\sim\left|\omega_{n}\right|^{\frac{2}{\alpha_{0}}-1}$
becomes divergent for $\alpha_{0}\geq2$ {[}see Appendix A{]}, the
critical lines are cut at this point what is depicted on phase diagrams
in Fig.\ref{Fig2}a and Fig.\ref{Fig3}. This boundary value $\alpha_{0}$
does not depend on magnetic field and on the geometry of the network.
It seems to be universal for different types of lattices without and
in presence of the external magnetic field. The system behaves as
if it were classical because of the big contribution to the action
Eq. (\ref{action}) which is generated by large values of the dissipation
parameter $\alpha_{0}$. Dissipation can correlate the phase on a
single island in different time and this correlation has the biggest
impact on global coherent state when $\alpha_{0}$ is greater then
critical value. Similar behaviour of the phase diagrams with critical
value of the dissipation parameter was predicted by several authors.\cite{chakravaraty,fisher1,wagenblast}
Theory developed by Chakravaraty et al.\cite{chakravaraty} reveals,
that phase transition takes place at point $\alpha_{\mathrm{crit}}=1/d$
where $\textrm{$d$}$ is the dimension of the system. The insulating
phase disappears for $\alpha_{0}>2$ in model considered by Wagenblast
et al.\cite{wagenblast}, the authors claim that dissipative processes
completely suppress the phase fluctuations. Kampf and Sch\"on\cite{kampf1}
using variational procedure showed that different mechanisms: Ohmic
and quasi-particle damping lead to different critical values of $\alpha_{0}$.
In magnetic field the dependence of critical temperature on several
ratio of the magnetic fluxes is periodic with period $1$ and symmetric
around $p/q=1/2$.\cite{kampf} Note that phase diagrams for different
lattice geometry and in presence of a magnetic field with effects
of the local dissipation has not been presented. %

\section{Summary}

We have studied the ground phase diagram two-dimensional Josephson
junction arrays in quantum regime by analytical methods using the
quantum-spherical approach with exactly evaluated density of states
for square and triangular lattice. For square lattice we could take
into consideration perpendicular magnetic field which was described
by rational magnetic flux $f=p/q$ for a number of values $1/q$.
Zero temperatures phase diagram indicates that for small values $\alpha_{0}$
quantum fluctuations destroy the long-range phase coherence. The arrays
can be in two phases: Mott-insulator phase and global coherent state.
We can travel between phases vary Coulomb energy or dissipation parameter.
When $\alpha_{0}$ is greater than critical value, the dissipation
suppresses quantum fluctuations and the array orders even for small
ratio $E_{J}/E_{C}$. Why dissipation can restore global coherent
state? If we imagine that we allow electrons tunnel between superconducting
arrays and metallic substrate, remaining Cooper's pairs will become
more mobile too. Therefore we will not be equal to set a number pair
on the island. Because superconducting phase $\phi$ and the number
of Cooper's pair $\hat{n}$ follow uncertainty relation $\left[\phi,\hat{n}\right]=2i$
then if we cannot say anything about quantity of Cooper's pair on
the island the phase is determined and in consequence we obtain global
coherent state. In case when system is in presence of the magnetic
field damping is stronger than in the absence. If we vary flux parameter
$f=p/q$ we could drive array into insulating state. Magnetic field
could affect the dissipation because it influences the resistance
of at least the coherent parts and thus changes the effective shunting.
For small values dissipation parameter $\alpha_{0}$ Josephson energy
in triangular lattice is damped less then for square lattice even
in presence magnetic field. We can notice that if the quantum fluctuations
of the phase superconducting order parameter are important, the dissipation
plays decisive role in constitute the onset of global superconductivity.

\appendix

\section{Low Temperature Properties of the corellator}

We write correlation function Eq. (\ref{correlation function}) in
form:\begin{equation}
\mathcal{W}\left(\tau\right)=\exp\left\{ -\frac{1}{\beta}\sum_{n\neq0}\frac{1-\cos\left(\omega_{n}\tau\right)}{\frac{1}{8E_{C}}\omega_{n}^{2}+\frac{\alpha_{0}}{2\pi}\left|\omega_{n}\right|}\right\} \end{equation}
and observe that the sum over $\omega_{n}$ is symmetric when we change
$\omega_{n}\rightarrow-\omega_{n}$. In that case (getting rid of
abs) we can write\begin{equation}
\mathcal{W}\left(\tau\right)=\exp\left\{ -\frac{2}{\beta}\sum_{n\geq1}\frac{1-\cos\left(\omega_{n}\tau\right)}{\frac{1}{8E_{C}}\omega_{n}^{2}+\frac{\alpha_{0}}{2\pi}\omega_{n}}\right\} \end{equation}
 noticing then sums over $n\geq1$ and $n\leq1$ are the same. Now
we are splitting above expression for two parts and neglecting the
$\omega_{n}^{2}$ in second contribution (in low temperatures more
important is part which contains $\omega_{n}$) we get:\begin{widetext}\begin{equation}
\mathcal{W}\left(\tau\right)=\exp\left[-\frac{2}{\beta}\sum_{n\geq1}\frac{1}{\frac{1}{8E_{C}}\omega_{n}^{2}+\frac{\alpha_{0}}{2\pi}\omega_{n}}\right]\exp\left[\frac{2}{\beta}\sum_{n\geq1}\frac{\cos\left(\omega_{n}\tau\right)}{\frac{\alpha_{0}}{2\pi}\omega_{n}}\right].\end{equation}
Evaluating the sums we can write:\begin{equation}
\mathcal{W}\left(\tau\right)=\exp\left[-\frac{1}{\alpha_{0}}\mathbf{H}\left(\frac{8\beta E_{C}}{\alpha_{0}}\right)\right]\exp\left\{ \frac{2}{\alpha_{0}}\log\left[2\sin\left(\frac{\pi}{\beta}\left|\tau\right|\right)\right]\right\} ,\end{equation}
and using asymptotic relation (valid for $\beta\rightarrow\infty$)
of the harmonic number function $\mathbf{H}\left(n\right)=\sum_{i=1}^{n}i^{-1}$:\begin{equation}
\mathbf{H}\left(a\beta\right)=\ln a\beta+\gamma+\frac{1}{2}a\beta-\frac{1}{12}\left(\frac{1}{a\beta}\right)^{2}+\frac{1}{120}\left(\frac{1}{a\beta}\right)^{4}+...\mathcal{O}\left(\frac{1}{\beta^{6}}\right),\end{equation}
where $\gamma$ is Euler's constant, finally we have found \begin{equation}
\mathcal{W}^{-1}\left(\tau\right)=\exp\left(\frac{\gamma}{\alpha_{0}}\right)\left(2\frac{\pi}{\beta}\left|\tau\right|\right)^{-\frac{2}{\alpha_{0}}}.\end{equation}
After Fourier transform correlator becomes:\begin{equation}
\mathcal{W}^{-1}\left(\omega_{n}\right)=\exp\left(\frac{\gamma}{\alpha_{0}}\right)\Gamma\left(1-\frac{2}{\alpha_{0}}\right)\sin\left(\frac{\pi}{\alpha_{0}}\right)\left|\omega_{n}\right|^{\frac{2}{\alpha_{0}}-1},\end{equation}
\end{widetext}besides the Euler gamma function $\Gamma\left(z\right)$
is defined by the integral:\cite{abramovitz} \begin{eqnarray}
\Gamma\left(z\right)=\int_{0}^{\infty}t^{z-1}e^{-t}dt & \textrm{for} & \Re z>0.\end{eqnarray}
and can be viewed as a generalization of the factorial function, valid
for complex arguments. Finally we can see that correlator \begin{equation}
\mathcal{W}^{-1}\left(\omega_{n}\right)\sim\left|\omega_{n}\right|^{\frac{2}{\alpha_{0}}-1}\end{equation}
at zero temperature diverges for $\alpha_{0}\geq2$.

\section{Density of States}

In the case of zero magnetic field the density of states for the square
two-dimensional lattice reads simply

\begin{equation}
\rho_{0}^{\square}\left(\xi\right)=\frac{1}{\pi^{2}}\mathbf{K}\left(\sqrt{1-\left(\frac{\xi}{2}\right)^{2}}\right)\Theta\left(1-\left|\frac{\xi}{2}\right|\right),\end{equation}
where \begin{equation}
\mathbf{K}\left(x\right)=\int_{0}^{\pi/2}\frac{d\phi}{\sqrt{1-x^{2}\sin^{2}\phi}},\end{equation}
 is the elliptic integral of the first kind\cite{abramovitz} and
the unit step function is defined by:\begin{equation}
\Theta\left(x\right)=\left\{ \begin{array}{ccc}
1 & \textrm{for} & x>0\\
0 & \textrm{for} & x\leq0\end{array}\right..\end{equation}

The procedure obtaining closed formulas for density of states in presence
of the magnetic field based on analytically solving Harper's equation.
The Harper's equation was studied intensively\cite{hasegawa1,gumbs,hong}
but only on numerical way and was missed analytic closed formulas
for density of states in presence of the magnetic field. Only expression
for case $p/q=1/2$ was received by Tan and Thouless and also was
formulated in terms of the elliptic integrals.\cite{tan} Therefore
the density of states for square lattice with the magnetic quantum
flux per plaquette for value $p/q=\frac{1}{2}$ reads:\cite{kopec2}\begin{equation}
\rho_{1/2}^{\square}\left(\xi\right)=\frac{\left|\xi\right|}{2}\rho_{0}^{\square}\left(\frac{\xi^{2}-4}{2}\right).\end{equation}
It is only one gap-less case instead of $p/q=0$.\begin{widetext}

Obtaining closed formulas for the next case $p/q=1/3$ was more difficult,
and we have got following expression\begin{eqnarray}
\rho_{1/3}^{\square}\left(\xi\right) & = & \frac{1}{4\sqrt{2}}\left|\xi^{2}-2\right|\sqrt{\xi^{2}-8}\rho_{0}^{\square}\left[\frac{1}{2}\xi\left(\xi^{2}-6\right)\right]\nonumber \\
 &  & \times\left\{ \left|\sec\left(\alpha+\frac{\pi}{2}\right)\right|\left[\Theta\left(\xi+1+\sqrt{3}\right)-\Theta\left(6-\xi^{2}\right)-\Theta\left(\xi-1-\sqrt{3}\right)\right]\right.\nonumber \\
 &  & +\sec\left(\alpha+\frac{\pi}{6}\right)\left[\Theta\left(\xi+\sqrt{6}\right)-\Theta\left(\xi+2\right)+\Theta\left(\xi\right)-\Theta\left(\xi+1-\sqrt{3}\right)\right]\nonumber \\
 &  & \left.+\sec\left(\alpha+-\frac{\pi}{6}\right)\left[\Theta\left(\xi-1+\sqrt{3}\right)-\Theta\left(\xi\right)+\Theta\left(\xi-2\right)\Theta\left(\xi-\sqrt{6}\right)\right]\right\} ,\end{eqnarray}
where\begin{equation}
\alpha=\frac{1}{3}\arctan\left(\frac{\sqrt{32-\xi^{2}\left(\xi^{2}-6\right)^{2}}}{\xi\left(\xi^{2}-6\right)}\right).\end{equation}
 For $p/q=1/4$ the expression for density of states is straightly
given by: \begin{eqnarray}
\rho_{1/4}^{\square}\left(\xi\right) & = & \frac{1}{2}\left|\xi^{2}-4\right|\rho_{0}^{\square}\left[\frac{1}{2}\left(\xi^{4}-8\xi^{2}+4\right)\right]\left\{ \sqrt{4+\left|\xi^{2}-4\right|}\left[\Theta\left(8-\xi^{2}\right)-\Theta\left(4+2\sqrt{2}-\xi^{2}\right)\right]\right.\nonumber \\
 &  & \left.+\sqrt{4-\left|\xi^{2}-4\right|}\Theta\left(4-2\sqrt{2}-\xi^{2}\right)\right\} .\end{eqnarray}
 Finally the most complicated case $p/q=1/6$ with six symmetric singularities
divide symmetrically on the positive and negative part of the axis
$\xi$:\begin{eqnarray}
\rho_{1/6}^{\square}\left(\xi\right) & = & \frac{1}{4\sqrt[4]{2}}\left|\xi^{4}-8\xi^{2}+8\right|\sqrt{\xi^{4}-8\xi^{2}-16}\rho_{0}^{\square}\left[\frac{1}{2}\left(\xi^{2}-6\right)\right]\nonumber \\
 &  & \times\left\{ \left|\sec\left(\alpha+\frac{\pi}{2}\right)\right|\left[\Theta\left(\xi+1+\sqrt{3}\right)-\Theta\left(6-\xi^{2}\right)-\Theta\left(\xi-1-\sqrt{3}\right)\right]\right.\nonumber \\
 &  & +\sec\left(\alpha+\frac{\pi}{6}\right)\left[\Theta\left(\xi+\sqrt{6}\right)-\Theta\left(\xi+2\right)+\Theta\left(\xi\right)-\Theta\left(\xi+1-\sqrt{3}\right)\right]\nonumber \\
 &  & \left.+\sec\left(\alpha+-\frac{\pi}{6}\right)\left[\Theta\left(\xi-1+\sqrt{3}\right)-\Theta\left(\xi\right)+\Theta\left(\xi-2\right)\Theta\left(\xi-\sqrt{6}\right)\right]\right\} ,\end{eqnarray}
where\begin{equation}
\alpha=\frac{1}{3}\arctan\left(\frac{\left|\left(\xi^{4}-8\xi^{2}+8\right)\sqrt{16+8\xi^{2}-\xi^{4}}\right|}{\xi^{6}-12\xi^{4}+24\xi^{2}+32}\right).\end{equation}
The density of states for triangular lattice with six nearest neighbors
we can write in form:\begin{equation}
\rho^{\bigtriangleup}\left(\xi\right)=\frac{2}{\pi^{2}\sqrt{\kappa_{0}}}\mathbf{K}\left(\sqrt{\frac{\kappa_{1}}{\kappa_{0}}}\right)\left[\Theta\left(\xi+\frac{3}{2}\right)-\Theta\left(\xi-3\right)\right],\end{equation}
where\begin{equation}
\kappa_{0}=\left(3+2\sqrt{3+2\xi}-\xi^{2}\right)\left[\Theta\left(\xi+\frac{3}{2}\right)-\Theta\left(\xi+1\right)\right]+4\sqrt{3+2\xi}\left[\Theta\left(\xi+1\right)-\Theta\left(\xi-3\right)\right],\end{equation}
\begin{equation}
\kappa_{1}=4\sqrt{3+2\xi}\left[\Theta\left(\xi+\frac{3}{2}\right)-\Theta\left(\xi+1\right)\right]+\left(3+2\sqrt{3+2\xi}-\xi^{2}\right)\left[\Theta\left(\xi+1\right)-\Theta\left(\xi-3\right)\right].\end{equation}
\end{widetext}


\begin{thebibliography}{10}
\bibitem{abeles}B. Abeles, Phys. Rev. B \textbf{15}, 2828 (1977). 
\bibitem{simanek1}E. \u{S}im\'anek, Solid State Commun. \textbf{31}, 419 (1979).
\bibitem{simanek2}E. \u{S}im\'anek, Phys. Rev. B \textbf{22}, 459 (1980); \textbf{23},
5762 (1982); \textbf{32}, R500 (1985).
\bibitem{doniach}S. Doniach, Phys. Rev. B \textbf{24}, 5063 (1981).
\bibitem{bradley}R. M. Bradley and S. Doniach, Phys. Rev. B \textbf{30}, 1138 (1984).
\bibitem{wood}D. M. Wood and D. Stroud, Phys. Rev. B \textbf{25}, 1600 (1982).
\bibitem{kopec1}T. K. Kope\'c and J. V. Jos\'e, Phys. Rev. B \textbf{63}, 064504
(2001).
\bibitem{jose}J. V. Jos\'e, Phys. Rev. B \textbf{29}, R2836 (1984); L. Jacobs,
J. V. Jos\'e and M. A. Novotny, Phys. Rev. Lett. \textbf{53}, 2177
(1984).
\bibitem{kopec2}T. K. Kope\'c and T. P. Polak, Phys. Rev. B \textbf{66}, 094517 (2002).
\bibitem{voss}R. F. Voss and R. A. Webb, Phys. Rev B \textbf{25}, R3446 (1982). 
\bibitem{wess}B. J. van Wees, H. S. J. van der Zant, and J. E. Mooij, Phys. Rev.
B \textbf{35}, R7291 (1987).
\bibitem{zant}H. S. J. van der Zant, W. J. Elion, L. J. Geerligs, and J. E. Mooij,
Phys. Rev. B \textbf{54}, 10081 (1996).
\bibitem{caputo}P. Caputo, M. V. Fistul, and A. V. Ustinov, Phys. Rev. B \textbf{63},
214510 (2001).
\bibitem{bruder}C. Bruder, R. Fazio, and G. Sch\"on, Phys. Rev. B \textbf{47}, 342
(1993); T. K. Kope\'c and J. V. Jos\'e, Phys. Rev. B \textbf{60},
7473 (1999), G. Grignani, A. Mattoni, P. Sodano, A. Trombettoni, Phys.
Rev. B \textbf{61}, 11676 (2000); W. A. Al-Saidi and D. Stroud, Phys.
Rev. B \textbf{67}, 024511 (2003); W. A. Al-Saidi and D. Stroud, Physica
C 402, 216, (2004).
\bibitem{caldeira}A. O. Caldeira and A. J. Leggett, Phys. Rev. Lett. \textbf{46}, 211
(1981); Ann. Phys. \textbf{149}, 374 (1983).
\bibitem{schon}G. Sch\"on, A. D. Zaikin, Phys. Rep. 198, 237 (1990).
\bibitem{ambegaokar}V. Ambegaokar, U. Eckern and G. Sch\"on, Phys. Rev. Lett. \textbf{48},
1745 (1982).
\bibitem{chakravaraty}S. Chakravarty, G. L. Ingold, S. Kivelson, and A. Luther, Phys. Rev.
Lett. \textbf{56}, 2303 (1986); S. Chakravarty, G. L. Ingold, S. Kivelson,
and G. Zim\'anyi, Phys. Rev. B \textbf{37}, 3283 (1988).
\bibitem{fisher}M. P. A. Fisher, Phys. Rev. Lett. \textbf{57}, 885 (1986); S. Chakravarty,
S. Kivelson, G. T. Zim\'anyi, and B. I. Halperin, Phys. Rev. B \textbf{35},
R7256 (1987).
\bibitem{simanek3}E. \u{S}im\'anek and R. Brown, Phys. Rev. B \textbf{34}, R3495 (1986).
\bibitem{fisher1}M. P. A. Fisher, Phys. Rev. \textbf{B} 36, 1917 (1987).
\bibitem{zwerger}W. Zwerger, J. Low Temp. Phys. \textbf{72}, 291 (1988).
\bibitem{panyukov}S. V. Panyukov, A. D. Zaikin, J. Low Temp. Phys. \textbf{75}, 365
(1989); \textbf{75}, 389 (1989).
\bibitem{chakravarty1}S. Chakravarty, G. L. Ingold, S. Kivelson, A. Luther, Phys. Rev. Lett.
\textbf{56}, 2303 (1986).
\bibitem{chakravarty2}S. Chakravarty, G. L. Ingold, S. Kivelson, G. Zim\'anyi, Phys. Rev.
B \textbf{37}, 3283 (1988).
\bibitem{falci}G. Falci, R. Fazio, G. Giaquinta, Europhys. Lett. \textbf{14}, 145
(1991).
\bibitem{cuccoli}A. Cuccoli, A. Fubini, V. Tognetti, R. Vaia, cond-mat=0002072.
\bibitem{eckern}U. Eckern, G. Sch\"on and V. Ambegaokar, Phys. Rev. B \textbf{30},
6419 (1984).
\bibitem{kampf1}A. Kampf, G. Sch\"on, Physica \textbf{152}, 239 (1988); A. Kampf,
G. Sch\"on, Phys. Rev. B \textbf{36}, 3651 (1987); E. \u{S}im\'anek
and R. Brown, Phys. Rev. B \textbf{34}, R3495 (1986).
\bibitem{choi}J. Choi and J. V. Jos\'e, Phys. Rev. Lett. \textbf{62}, 1904 (1989).
\bibitem{kampf}A. Kampf and G. Sch\"on, Phys. Rev. B \textbf{37}, R5954 (1988).
\bibitem{kim}S. Kim and M. Y. Choi, Phys. Rev. \textbf{B} 42, 80 (1990).
\bibitem{beck}H. Beck, Phys. Rev. \textbf{B} 49, 6153 (1994); S. E. Korshunov, Phys.
Rev. \textbf{B} 50, 13616 (1994); K. H. Wagenblast, A. van Otterlo,
G. Sch\"on, G. T. Zim\'anyi, Phys. Rev. Lett. \textbf{79}, 2730
(1997).
\bibitem{geerlings}L. J. Geerligs, M. Peters, L. E. M. de Groot, A. Verbruggen, and J.
E. Mooij, Phys. Rev. Lett. \textbf{63}, 326 (1989); A. F. Hebard and
M. A. Paalanen, Phys. Rev. Lett. \textbf{65}, 927 (1990); A. J. Rimberg,
T. R. Ho, \c{C}. Kurdak, J. Clarke, K. L. Campman, A. C. Gossard,
Phys. Rev. Lett. \textbf{78}, 2632 (1997).
\bibitem{valles}J. M. Valles, Jr. R. C. Dynes, and J. P. Garno, Phys. Rev. Lett. \textbf{69},
3567 (1992); 
\bibitem{yazdani}A. Yazdani and A. Kapitulnik, Phys. Rev. Lett. \textbf{74}, 3037 (1995).
\bibitem{fazio}R. Fazio and G. Sch\"on, Phys. Rev. B \textbf{43}, 5307 (1991).
\bibitem{homes}C. C. Homes, S. V. Dordevic, M. Strongin, D. A. Bonn, Ruixing Liang,
W. N. Hardy, Seiki Komiya, Yoichi Ando, G. Yu, N. Kaneko, X. Zhao,
M. Greven, D. N. Basov, T. Timusk, Nature, \textbf{430}, 539 - 541
(2004).
\bibitem{zurek}W. H. Zurek, Phys. Today, \textbf{44}, 36 (1991); D. Braun, F. Haake,
W. T. Strunz, Phys. Rev. Lett. \textbf{86}, 2913 (2001).
\bibitem{han}S. Han, Y. Yu, X. Chu, S. Chu, Z. Wang, Science, \textbf{293}, 1457
(2001).
\bibitem{shih}W. Y. Shih and D. Stroud, Phys. Rev. B \textbf{30}, R6774 \textbf{(}1984\textbf{).}
\bibitem{granato}E. Granato and J. M. Kosterlitz, Phys. Rev. Lett. \textbf{65}, 1267
(1990).
\bibitem{benz}S. P. Benz, M. S. Rzchowski, M. Tinkham and C. J. Lobb, Phys. Rev.
Lett. \textbf{64}, 693 (1990).
\bibitem{abramovitz}M. Abramovitz and I. Stegun, Handbook of Mathematical Functions (Dover,
New York, 1970). 
\bibitem{berlin}T. H. Berlin and M. Kac, Phys. Rev. B \textbf{86}, 821 (1952); H.
E. Stanley, Phys. Rev. B \textbf{176}, 718 (1968). 
\bibitem{joyce}G. S. Joyce Phys. Rev. B \textbf{146}, 349 (1966).
\bibitem{qsa}The spherical model in general reveals, in a simplified way, the failure
of mean field theory below the upper critical dimension. For example,
in the classical spherical model, the critical exponent of the correlation
length is $\nu=1/\left(d-2\right)$ which differs from the mean field
value $\nu=1/2$, which is dimensionality independent. Spherical model
also predicts that critical temperature is vanishing to zero if dimensionality
of the system $d$ is less then two, which is not suprising according
to Mermin-Wagner theorem.
\bibitem{wagenblast}K. H. Wagenblast, A. van Otterlo, G. Sch\"on and G. T. Zim\'anyi,
Phys. Rev. Lett. \textbf{78}, 1779 (1997).
\bibitem{kopec3}T. K. Kope\'c, J. V. Jos\'e, Phys. Rev. B \textbf{60}, 7473 (1999).
\bibitem{harper}P. G. Harper, Proc. Phys. Soc. London Sect. A \textbf{68}, 674 (1955).
\bibitem{hasegawa}Y. Hasegawa, P. Lederer, T. M. Rice, P. B. Wiegmann, Phys. Rev. Lett.
\textbf{63}, 907 (1984).
\bibitem{hasegawa1}Y. Hasegawa, Y. Hatsugai, M. Kohmoto and G. Montambaux, Phys. Rev.
B \textbf{41}, 9174 (1990).
\bibitem{gumbs}G. Gumbs and P. Fekete, Phys. Rev. B \textbf{56}, 3787 (1997).
\bibitem{hong}S. P. Hong and S.-H. S. Salk, Phys. Rev. B \textbf{60}, 9550 (1999).
\bibitem{tan}Y. Tan and D. J. Thouless, Phys. Rev. B \textbf{46}, 2985 (1992).
\end{thebibliography}
\end{document}